\documentclass[aps,prl,twocolumn,showpacs]{revtex4}

\usepackage{amsfonts}
\usepackage{amsbsy}
\usepackage{amsmath}

\usepackage{amsmath,amssymb,bm}
\usepackage{graphicx}
\usepackage{color}

\def\be{\begin{equation}}
\def\ee{\end{equation}}
\def\ba{\begin{eqnarray}}
\def\ea{\end{eqnarray}}

\begin{document}

\title{
Photo-induced separation of chiral isomers in a classical buffer gas}
\author{B. Spivak}
\affiliation{Physics Department of the University of Washington,
Seattle WA 98195}
\author{A. V. Andreev} \affiliation{Physics Department of the University of
Washington, Seattle WA 98195}
\begin{abstract}
We develop a theory of photo-induced drift of chiral molecules or
small particles in classical buffer gases. In the absence of a
magnetic field there exists a flux of chiral molecules, provided the
electromagnetic field is circularly polarized. It has opposite signs
for different chiral isomers. In the presence of a magnetic field
the flux can be also induced by a linearly polarized (or
unpolarized) electromagnetic field. The magnitude of the flux is not
proportional to either linear or orbital momentum of the
electromagnetic field.
\end{abstract}
\date{August 22, 2008}
\pacs{37.10.Vz, 42.62.Be,  47.45.Ab} \maketitle

In this article we develop a theory of photo-induced transport of
chiral molecules (or small chiral particles) diluted in a classical
buffer gas. This effect can be important for separation of different
isomers of chiral molecules and for the theory of optically actuated
molecular motors (see for example
\cite{Saha,Wiel,Koumura,Kelly,Polland}).

Consider a solution of molecules of definite chirality in a buffer
gas in the presence of an electromagnetic radiation with frequency
$\omega$ and polarization tensor $P_{ij}= \langle E_i E_j^*\rangle$,
where $\mathbf{E}$ is electric field and $\langle \ldots \rangle$
denotes time averaging. We define the chiral current $\mathbf{j}_c$
as the part of molecular current that changes sign when the
molecules are replaced by stereo-isomers of opposite chirality. We
study effects whose magnitudes are not proportional to either linear
or angular momentum of light, i.e. we work in the approximation
where no force or torque is exerted by the electromagnetic radiation
on the medium. This approximation is valid as long as the thermal
angular momentum of the molecules is larger than $\hbar$. In this
case in the absence of a magnetic field, by symmetry, the chiral
flux density is nonzero only if the electromagnetic field is
circularly polarized,
\begin{equation}\label{eq:1}
 j_{c,k}=  \alpha_{c} \epsilon_{ijk} P_{ij}.
\end{equation}
Here $\epsilon_{ijk}$ is the antisymmetric tensor,  and the
coefficient $\alpha_{c}$ has opposite signs for molecules of
different chirality. It is clear from Eq.~(\ref{eq:1}) that when the
direction of circular polarization is reversed the chiral current
changes sign.

Although the photo-induced flux of chiral molecules exists both in
gases and liquids the microscopic mechanisms of the effect in these
two cases are different. In this article we focus on the case of
gases, where all relevant mean free paths (for both buffer gas and
chiral molecules) are bigger than sizes of the molecules. In this
case the effect can be analyzed using the Boltzmann kinetic
equation.

We assume that  optical transitions take place between the ground state and an
exited state of the chiral molecules denoted by indices $g$ and $e$ respectively,
the energy difference between them being $\hbar \omega_{0}$.

\begin{figure}[ptb]
\includegraphics[width=8.0cm]{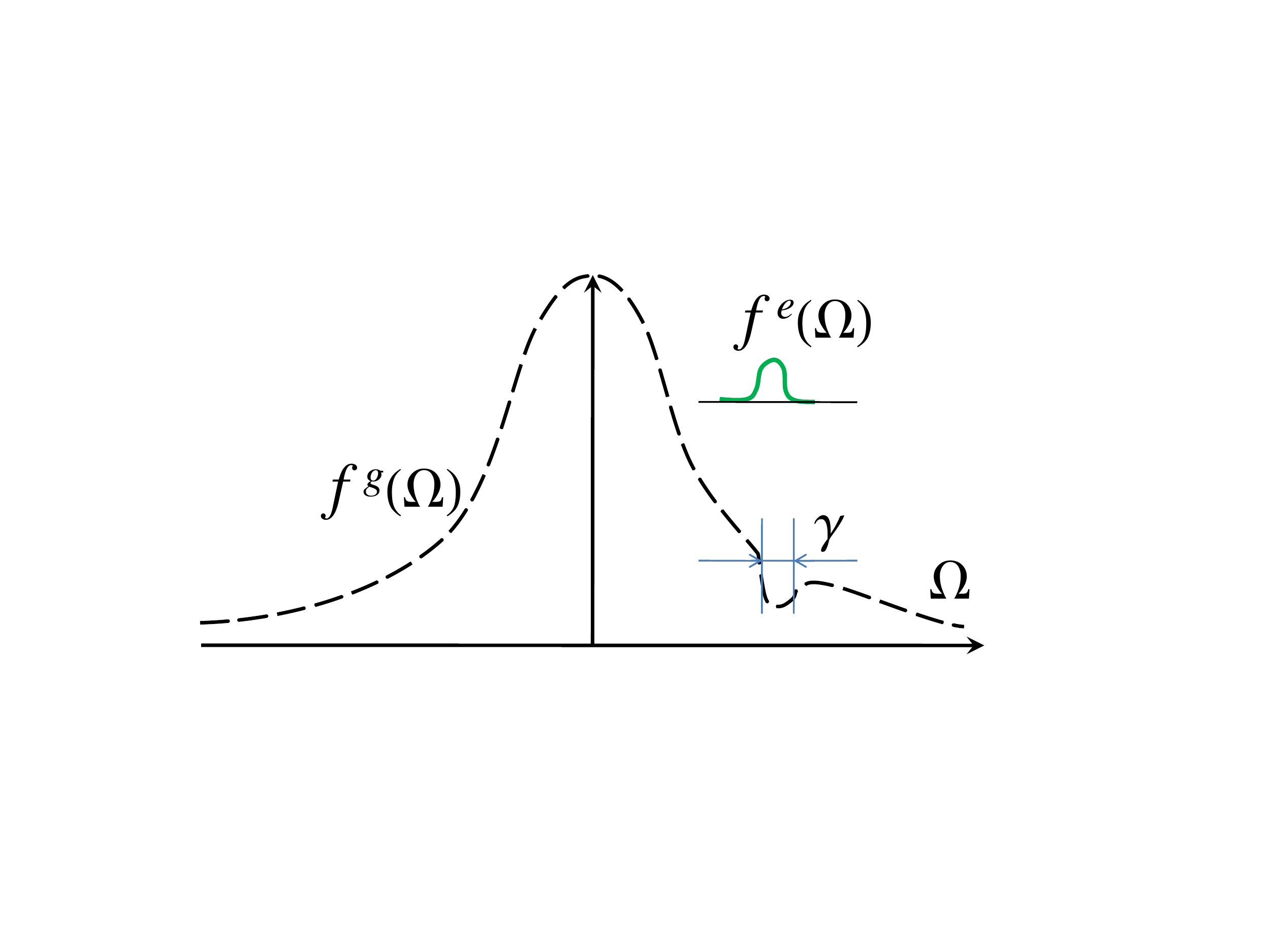}
\caption{(a) Schematic drawing of the dependence of the distribution
function of chiral molecules on their angular velocity along the
wavevector of light. } \label{fig:1}
\end{figure}

Let us first illustrate the origin of the effect by considering a
toy model in which the chiral molecules are constrained to rotate
about a single axis which coincides with the direction of
propagation of light. The frequency of the circularly polarized
radiation in the reference frame co-rotating with the molecule is
shifted by the rotation frequency of the molecule, $\Omega$. As a
result the probabilities of light absorption by molecules rotating
clockwise and counterclockwise are different. Consequently the
distribution function of the molecules becomes asymmetric with
respect to the angular velocity: a hole will occur in the
distribution function of the ground state molecules $f^{g}(\Omega)$
at certain values of the angular velocity, while a peak will occur
in the exited state distribution $f^e (\Omega)$(see
Fig.~\ref{fig:1}). Thus the populations of molecules in the excited
and ground states acquire opposite angular momenta. A fundamental
property of collisions of the chiral molecules with the buffer gas
is that they transfer rotation of chiral molecules into translation.
Since the scattering cross-sections for the excited and ground
states are generally different this will result in the net flux of
the chiral molecules.  Momentum conservation during collisions
implies that a net momentum in the direction opposite to the flux is
imparted to the buffer gas. In some aspects this  effect is similar
to the light-induced drift of atoms considered in
Refs.~\onlinecite{Shalagin,Dykhne,Woerdman,Kuscer}.

We now turn a description of the effect in the framework of the
Boltzmann kinetic equation. Due to the rapid mixing of rotational
degrees of freedom of the molecules, their distribution function
depends only on the center of mass coordinate $\mathbf{r}$, time
$t$, and the set of quantities $\Gamma$ that are conserved during
free motion~\cite{LandauKin}.  Below we consider a spatially-uniform
situation and assume that the chiral particles scatter only from
non-chiral buffer gas, whose distribution function is in
equilibrium. Then  the Boltzmann kinetic equation has the form
\begin{equation}
\label{eq:Boltzmann}
 \frac{\partial \tilde{ f}^{a}( \Gamma
)}{\partial t}= I^{a}\{ \tilde f(\Gamma)\}+ I_{ph}^{a}(\Gamma).
\end{equation}
Here $\tilde f^a(\Gamma)$ is the nonequilibrium part of the
distribution function, and  $a=g,e$ labels the ground or exited
states of the chiral molecules. Denoting the scattering probability
of a molecule from state $a,\Gamma$ to $ a', \Gamma' $ by
$w^{a'a}(\Gamma';\Gamma)$ we write the scattering integrals in the
excited and ground state due to collisions with the buffer gas as
\begin{subequations}
\label{eq:col_int}
\begin{eqnarray}
\label{eq:col_int_a} I^{g}\{ \tilde f(\Gamma)\}&=& \int w^{gg}(\Gamma;\Gamma')
\tilde f^g(\Gamma ') \,
d\Gamma ' \nonumber \\
&&-\tilde f^{g}(\Gamma) \int w^{gg}(\Gamma';\Gamma) \, d
\Gamma'  \nonumber \\
&&+ \int w^{ge}(\Gamma;\Gamma') \tilde f^e(\Gamma') d \Gamma',\\
I^{e}\{ \tilde f(\Gamma)\}&=& -\tilde f^{e}(\Gamma) \int w^{ge}(\Gamma';\Gamma)\,
d \Gamma'. \label{eq:col_int_b}
\end{eqnarray}
\end{subequations}
To linear order in the light intensity $\mathcal{I}$ the transition
rates between the ground and exited states due to light absorption
are described  by
\begin{subequations}
\label{eq:I_absorption}
\begin{eqnarray}
\label{eq:I_absorption_a} I_{ph}^{g}(\Gamma)&=& -f_{0}(\Gamma)\,
\Upsilon(\Gamma)\, \mathcal{I},\\
\label{eq:I_absorption_b} I_{ph}^{e}(\Gamma)&=& f_{0}(\Gamma) \,
\Upsilon(\Gamma)\, \mathcal{I},
\end{eqnarray}
\end{subequations}
where $f_0(\Gamma)$ is the equilibrium distribution function of the
molecules in the ground state (we assume that the excited state is
not thermally populated), and $\Upsilon (\Gamma)$ is the probability
of light absorption. The latter is determined by the time dependence
of the electric field in the reference frame of the molecule, and in
general depends both on the molecule velocity (due to the Doppler
effect) and its angular velocity. The rotational frequency shift is
of order of the angular velocity of rotation, $\sim v_T/d$, where
$v_T$ is the thermal speed of the molecule and $d$ is its
characteristic size. The Doppler frequency shift is much smaller, of
order $v_T/\lambda$, where  and $\lambda$ is the wavelength of
light. We assume that the mismatch between the radiation frequency
$\omega$  and the resonant frequency $\omega_0$ is of the order of
the thermal rotational shift, i.e. $|\omega -\omega_0| \sim v_T/d$,
and that the absorption line width $\gamma$ is smaller than the
rotational broadening $\gamma \ll v_T/ d$. In this regime the
Doppler effect can be neglected.

Thus we take the absorption coefficient $\Upsilon(\Gamma)$ to be
independent of the linear momentum of the chiral molecules. Since we
work in the dipole approximation, in which $\Upsilon(\Gamma)$ is
insensitive to the molecule chirality it is invariant under
inversion. In other words $I^a_{ph}(\Gamma)=I^a_{ph}(\Gamma^P)$,
where the set of coordinates $\Gamma^P$ is obtained from $\Gamma$ by
inversion. If the molecules were non-chiral the scattering
probabilities would be invariant under inversion, i.e.
\begin{equation}
w(\Gamma';\Gamma)=w(\Gamma'^{P};\Gamma^{P}). \label{invsymmetry}
\end{equation}
In this case the nonequilibrium part of the distribution function
$\tilde f (\Gamma) $ would be a symmetric function of momentum,
$\tilde f (\Gamma) =\tilde f (\Gamma^P)$. This would nullify the
chiral current
\begin{equation}\label{eq:chiral_current}
{\bf j}_c=\sum_{ a}  \int d \Gamma \, \mathbf{ v}\,  \tilde f^{a}
(\Gamma),
\end{equation}
where $\mathbf{v}$ is the molecule velocity. For chiral molecules,
in addition to inverting velocity, the inversion symmetry changes
the molecule chirality. In this situation the equality
Eq.~(\ref{invsymmetry}) is not valid, and the nonequilibrium
distribution $\tilde f (\Gamma) $ contains a part that is odd in
momentum and the chiral current Eq.~(\ref{eq:chiral_current})
generally does not vanish.

In order to complete the description of kinetics of the molecules in
the presence of radiation we need to obtain the dependence of the
absorbtion probability on the phase space coordinates,
$\Upsilon(\Gamma)$. In the dipole approximation $\Upsilon(\Gamma)$
can be expressed in terms of the three frequency dependent
absorbtion cross-section, $\sigma_i(\omega)$, for the different
light polarizations, by transforming the electric field from the lab
frame to the reference frame co-rotating with the molecule. This
problem can be solved for the general case, where molecule rotation
is described by an asymmetric top~\cite{LandauM}. However, in order
to simplify the discussion and illustrate the essential phisics we
consider the case where the molecule rotation is modeled by a
symmetric top.

\begin{figure}[ptb]
\includegraphics[width=8.0cm]{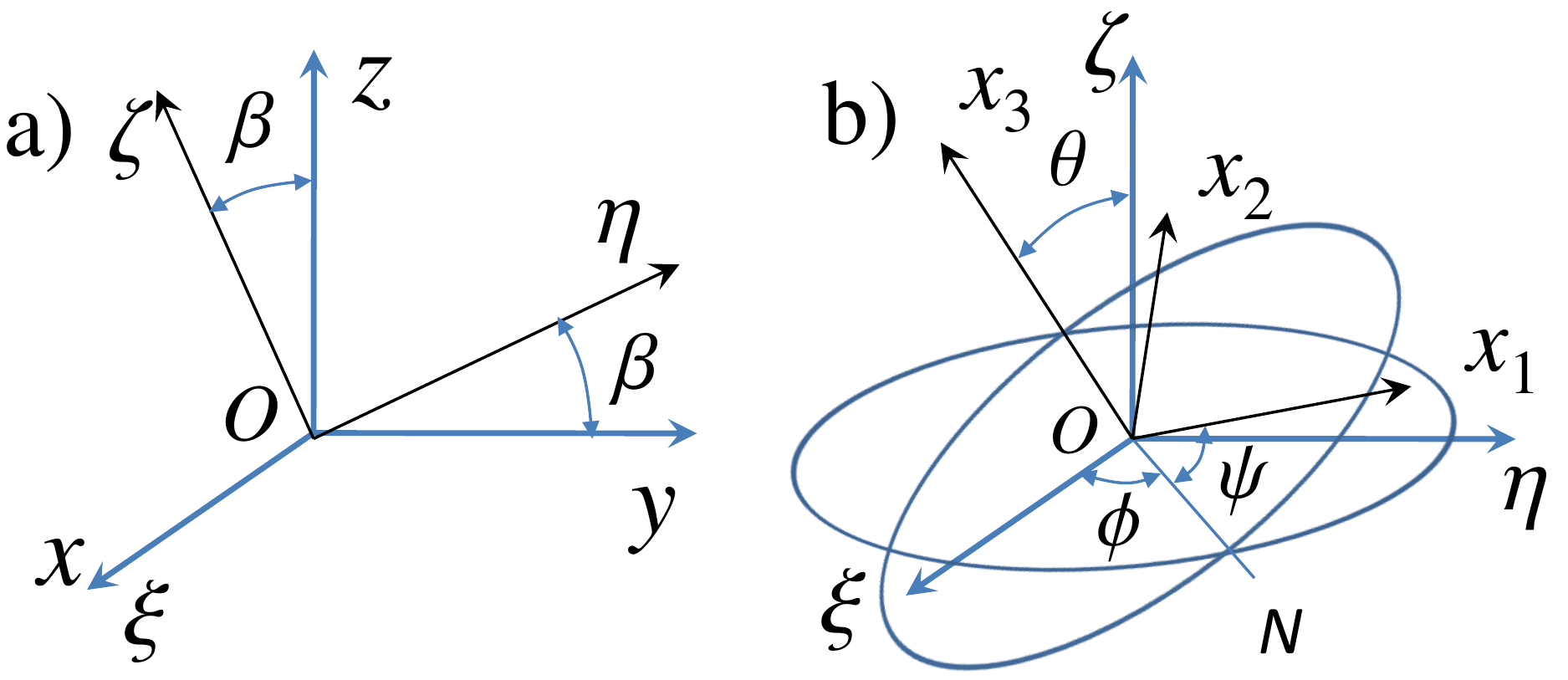}
\caption{(a) Rotation by the angle $\beta$ from the $xyz$ frame to
the $\xi\eta \zeta$ frame. (b) Illustration of the Euler angles
$\theta$, $\phi$, $\psi$ rotating the $\xi\eta \zeta$ frame to the
$x_1 x_2 x_3$ frame co-rotating with the molecule.} \label{fig:2}
\end{figure}

In this case the set of conserved quantities consists of the linear
momentum $\mathbf{p}$, angular momentum $\mathbf{M}$, and the angle
$\theta$ between the molecular axis and $\mathbf{M}$, i.e.
$\Gamma=\{\mathbf{p},\mathbf{M},\theta \}$ (we introduce the Euler
angles $\phi$, $\theta$, and $\psi$, as shown in Fig.~\ref{fig:2}
b))\footnote{As was mentioned, because of the fast change of the
angles $\psi$ and $\phi$ the distribution function of chiral
molecules depends only on the conserved angle $\theta$ between the
molecule axis and the angular momentum.}. In the case of circularly
polarized light the distribution function depends not on all three
components of the angular momentum $\mathbf{M}$ but only on its
magnitude $M$ and the angle $\beta$ it makes with the direction of
light propagation ($z$ axis in our notations, see Fig.~\ref{fig:2}
a)). Thus the distribution function $\tilde f$ depends only on ${\bf
p}, M, \theta, \beta$. Furthermore, it is convenient to express $M$
and $\theta$ in terms of the components of the angular momentum
parallel, $M_{3}=M\cos \theta $, and perpendicular, $M_{\perp}=M\sin
\theta$, to the symmetry axis of the molecule. Indeed the rotational
energy of a symmetric top is most naturally expressed in these
variables;
$\varepsilon(\Gamma)=\frac{M_{3}^{2}}{2I_{3}}+\frac{M_{\perp}^{2}}{2I_{\perp}}$.
Therefore below we write the equilibrium distribution function and
the kinetic equation in the variables $\Gamma =\{ {\bf p}, M_3,
M_\perp, \beta \}$,  with the integration measure  $d \Gamma \sim
d^3 p \, d \cos \beta \,  M_\perp d M_\perp d M_3$.

We assume that the excited state is non-degenerate. In this case,
for a symmetric top molecule and in the dipole approximation, only
the electric field polarized along the symmetry axis, $x_3$, of the
molecule is absorbed. Furthermore, the absorption cross-section for
the right and left stereo-isomers are identical and will be denoted
by $\sigma(\omega)$.

In order to express $\Upsilon(\Gamma)$ in terms of $\sigma(\omega)$
we need to transform the electromagnetic field $\mathbf{E}$ from the
stationary frame to the frame rotating with the molecule. To this
end we introduce three reference frames as illustrated in
Fig.~\ref{fig:2}. The $xyz$ frame is the lab frame such that the
$z$-axis corresponds to the direction of light propagation. The
light is assumed circularly polarized with the polarization vector
$\mathbf{E}=E_0( \hat x + i \hat y)/\sqrt{2}$, where the hat above a
symbol is used to denote a unit vector along the corresponding axis.
The $\xi \eta\zeta$ frame is also stationary but the $\zeta$ axis
points along the constant angular momentum $\mathbf{M}$ of a given
molecule and forms an angle $\beta$ with the $z$-axis. The $\xi$ and
the $x$ axes coincide. In terms of the basis vectors $\hat{\xi}$,
$\hat{ \eta}$, $\hat{\zeta}$ of this frame the electric field is
written as
\begin{equation}\label{eq:E_m_def}
    \mathbf{E}=E_+\frac{\hat{\xi}+i\hat{\eta}}{\sqrt{2}}+ E_\zeta \,
    \hat{\zeta}+E_-\frac{\hat{\xi}-i\hat{\eta}}{\sqrt{2}}, \nonumber
\end{equation}
with the components expressed in terms of $E_0$ as
\begin{equation}\label{eq:E_m}
    \left(
      \begin{array}{ccc}
        E_+, &
        E_\zeta, &
        E_-
      \end{array}
    \right)=E_0\left(
                 \begin{array}{ccc}
                   \frac{1+ \cos\beta }{2}, &
                    \frac{ -i \sin \beta}{\sqrt{2}}, &
                   \frac{1-\cos\beta}{2}
                 \end{array}
               \right).
\end{equation}
Next we transform the electric field from the fixed frame $\xi \eta\zeta$ to the
frame $x_1x_2x_3$ which rotates with the molecule so that the $x_3$-axis always
point along the symmetry axis of the molecule. The rotation from the $\xi
\eta\zeta$ frame to the $x_1x_2x_3$ frame is described by the Euler angles $\phi,
\theta, \psi$.

The components of the electric field in the rotating frame, $
\mathbf{E}=\tilde{E}_+\frac{\hat{x}_1+i\hat{x}_2}{\sqrt{2}}+ \tilde{E}_0 \,
\hat{x}_3+\tilde{E}_-\frac{\hat{x}_1- i\hat{x}_2}{\sqrt{2}}$, are given by
\begin{equation}\label{eq:rotation}
    \left(
      \begin{array}{c}
        \tilde{E}_+ e^{-i\psi}\\
        \tilde{E}_0 \\
        \tilde{E}_-e^{i\psi} \\
      \end{array}
    \right)=
    \left(
      \begin{array}{ccc}
        \frac{1+\cos\theta}{2} & \frac{-i \sin\theta}{\sqrt{2}} & \frac{1-\cos\theta}{2} \\
        \frac{- i \sin\theta}{\sqrt{2}} & \cos\theta & \frac{i \sin\theta}{\sqrt{2}} \\
        \frac{1-\cos\theta}{2}  & \frac{i \sin\theta}{\sqrt{2}} & \frac{1+ \cos\theta }{2} \\
      \end{array}
    \right)
    \left(
      \begin{array}{c}
        E_+ e^{i\phi}\\
        E_\zeta \\
        E_-e^{-i\phi} \\
      \end{array}
    \right).
\end{equation}
For a symmetric top molecule the angles $\phi$ and $\psi$ change
linearly with time while the angles $\beta$ and $\theta$ do not
change during free propagation. Thus it is clear from
Eq.~(\ref{eq:rotation}) that in the reference frame co-moving with
the molecule each component of the electric field, $\tilde{E}_\pm$,
and $\tilde{E}_0$, consists of three monochromatic waves with
frequencies of the form
\begin{equation}\label{eq:omega_pq}
    \omega_{pq} \equiv \omega - p \dot\psi - q \dot\phi,
\end{equation}
where  $p$ and $q$ can take the values $p,q= 0,\pm 1$. Since only
the field $\tilde{E}_0$ along the molecule axis $x_3$ is absorbed,
we can write  the absorption probability using
Eq.~(\ref{eq:rotation}) as
\begin{eqnarray}\label{eq:Upsilon_rot}
   \Upsilon(\Gamma)
     &=& \frac{1}{\hbar \omega} \left[
    \frac{\sin^2\theta}{8}\sum_{q=\pm 1}(1+q \cos \beta)^2 \sigma (\omega_{0\,
    q}) \right. \nonumber \\
    &&\left.
+ \, \frac{\cos^2\theta \sin^2 \beta}{2} \, \sigma (\omega_{0\, 0})
\right].
\end{eqnarray}
The frequencies $\omega_{pq}$ can be expressed in terms of the
variables $M_3$ and $M_\perp$, $M=\sqrt{M_\perp^2
  +M_3^2}$, using the kinematic
relations
\begin{equation}\label{eq:psi_dot_L}
  \dot\psi = \frac{M_3}{I_3}-\frac{M_3}{I_\perp} , \quad
  \dot\phi = \frac{M}{I_\perp},\quad
  \cos\theta = \frac{M_3}{M}.
\end{equation}

Equations (\ref{eq:Upsilon_rot}), (\ref{eq:omega_pq}) and (\ref{eq:psi_dot_L})
define the absorption probability as a function of the phase space coordinates
$\Gamma=\{ \mathbf{p}, M_3, M_\perp, \beta \}$. This probability is peaked when
the resonance condition $p \dot\psi + q \dot\phi =\omega -\omega_0$  is
satisfied.  In terms of the variables $M_3, M_\perp$ the resonance condition is
\[
 pM_3\left(\frac{I_\perp}{I_3}-1 \right)+q \sqrt{M_3^2+M_\perp^2}=
 I_\perp (\omega -\omega_0).
\]

We note that the light absorbtion probability
Eq.~(\ref{eq:Upsilon_rot}) contains a term that is linear in
$\cos\beta$. The coresponding optical transitions produce a
nonequilibrium distribution function that is odd in the component of
the angular momentum along the light propagation direction, $M_z$.
Subsequent collisions of the chiral molecules with the buffer gas
result in a net flux of the molecules along the $z$-axis.

In order to obtain an estimate for the magnitude of the effect we
consider the simplest case when the relaxation time $\tau_{eg}$ from
excited to ground state is the shortest time in the problem. Here $
1/\tau^{ge}(\Gamma)=\int d \Gamma' w^{ge}(\Gamma,\Gamma')$. Then in
a stationary state we get from Eqs.~(\ref{eq:Boltzmann}),
(\ref{eq:col_int_b}) and (\ref{eq:I_absorption_b})
\begin{equation}
\tilde f^{e}(\Gamma)= \tau^{ge}(\Gamma)\mathcal{I}\Upsilon(\Gamma)
f^{g}_{0}(\Gamma),
\end{equation}
while the Boltzmann equation for the ground state distribution
function becomes
\begin{eqnarray}
&&\tilde f^{g}(\Gamma)\int d\Gamma ' w^{gg}(\Gamma ';\Gamma ) - \int
d\Gamma ' w^{gg}(\Gamma; \Gamma ')\tilde f^{g}(\Gamma ')
 =  \nonumber \\
 &&\mathcal{I} \left[ -\Upsilon(\Gamma) f_0^g (\Gamma) +
 \int d\Gamma ' w^{ge} (\Gamma; \Gamma') \tau^{ge} (\Gamma ')
 \Upsilon(\Gamma ') f_0^g (\Gamma ')  \right]. \nonumber
\end{eqnarray}
In our approximation the molecules in the excited state do not
contribute to the total flux of chirality, i.e.
\[
{\bf j}_c=\int {\bf v} \tilde f^{g}(\Gamma) d \Gamma.
\]
Thus we are interested only in the anisotropic in ${\bf v}$ part of
$\tilde f^{g}(\Gamma)$,  which can be estimated as
\begin{eqnarray}\label{eq:jGasH=0}
{\bf j}_c &\sim & \tau_{p}^{gg} \mathcal{I}\int d \Gamma d \Gamma'  {\bf v}
\Upsilon(\Gamma ' ) f_{0}^{g} (\Gamma ') \times \nonumber \\
&&[ w^{ge}(\Gamma, \Gamma') \tau^{ge} (\Gamma ') -w^{gg}(\Gamma,\Gamma')
\tau^{gg}(\Gamma ')].
\end{eqnarray}
Here we introduced the relaxation time for scattering of chiral
molecules in the ground state, $\tau^{gg}$, and the momentum
relaxation time, $\tau^{gg}_{p}$.

A rough estimate for the chiral current Eq.~(\ref{eq:jGasH=0}) can be written as
\begin{eqnarray}
j_c\sim v_{T}n_{c}(B\tau_{p}) (\delta W^{ge} \tau^{ge}-\delta W^{gg}
\tau^{gg})\approx \nonumber \\ v_{T}\frac{I\tau_{p}}{\xi \hbar
\omega}(\delta W^{ge} \tau^{ge}-\delta W^{gg} \tau^{gg})
\label{estimate}
\end{eqnarray}
Here $n_{c}$ is the concentration of the chiral molecules, $v_{T}$
is their thermal velocity, $\xi$ is the absorption length of the
light, and $B\sim n_c^{-1} \mathcal{I} \int d\Gamma \Upsilon(\Gamma)
f_0(\Gamma) $ is the rate of the optical transitions in an
individual molecule. Here we introduced the ``chiral part'' of the
scattering rate as $\delta W^{ga}\sim ( B n_c v_T)^{-1}\int d\Gamma
d\Gamma' \hat z \cdot \mathbf{v}
[w^{ga}(\Gamma;\Gamma')-w^{ga}(\Gamma^P;\Gamma'^P)] f_0^g (\Gamma ')
\Upsilon (\Gamma ') \mathcal{I}$, with $a=e/g$. Thus $\delta
W^{ga}\tau^{ga}$ is a dimensionless measure of the degree of
chirality of the molecules (we are not aware of any studies of this
quantity). Equation (\ref{estimate}) holds at small enough
intensities of radiation, when $\tilde{ f} \ll f_{0}$. At large
intensities the magnitude of the effect saturates, and  the maximum
flux of chiral molecules is of order
\begin{equation}
j^{(max)} \sim v_{T}n_{c}\left(\frac{\gamma}{T}\right) (\delta
W^{ge}-\delta W^{gg}) \tau^{gg}_{p}.
\end{equation}
We note that relative signs  of $\delta W^{ge}$ and $\delta W^{gg}$, as well as
the  sign of the chiral current, are arbitrary and depend on the structure of the
molecules.

The effects considered above are direct analogues of photo-galvanic
effects in non-centrosymmetric crystals. There are two types of
photo-galvanic effects. The first one is associated with a transfer
of momentum from light to electrons, and its magnitude is
proportional to the momentum of light. The analogue of this effect,
which is associated with transfer of angular momentum of
electromagnetic field to molecules in classical gases and liquids
has been considered in
Refs.~\onlinecite{Baranova78,Grier,Grier1,Grier2}. The second type
of the photo-galvanic effects exist only in non-centrosymmetric
crystals and is not proportional to the momentum of light
\cite{Belinicher,fridkin,ivchenko,IvchenkoPikus}. Since the momentum
of light is small the magnitude of this effect is generally bigger
than that of the first one. The photo-induced drift of chiral
molecules diluted in a buffer gas considered here, is an analogue of
the second type of photo-galvanic effect.

There exists also an analogue of the linear photo-galvanic
effect~\cite{Belinicher,fridkin,ivchenko,IvchenkoPikus}. Namely, in the presence
of a magnetic field a flux of chiral particles exists even if the electromagnetic
field is linearly polarized or unpolarized,
\begin{equation}
j_{c,k}=  \tilde{\alpha}_{c}  H_{k} P_{jj} +\tilde{\beta}_c ( P_{kj}+ P_{jk})H_j.
\nonumber
\end{equation}
The photo-induced chiral drift also exists when chiral molecules are
dissolved in a liquid. Photo-induced propulsion of molecules in this
situation can be described in the framework of hydrodynamics. These
two aspects of the phenomenon will be considered in a separate
publication.

We acknowledge useful discussions with  A. Grosberg, E. Ivchenko, B. Shklovskii,
D. Son and G. Tartakovski. The work was supported by  NSF grant DMR-0704151 and
DOE grant DE-FG02-07ER46452.

\end{document}